\documentstyle[11pt]{article}
\newcommand{\blankline}{\vskip .3cm}
\newcommand{\f}{\begin{equation}}
\newcommand{\ff}{\end{equation}}
\begin{document}
\centerline{\LARGE The exceptional Jordan algebra and the matrix string}
\blankline
\rm
\centerline{Lee Smolin}
\blankline
\centerline{\it  Center for Gravitational Physics and Geometry}
\centerline{\it Department of Physics, The Pennsylvania State University}
\centerline{\it University Park, PA, USA 16802 \ \ and }
\centerline{\it The Blackett Laboratory,Imperial College of 
Science, Technology and Medicine }
\centerline{\it South Kensington, London SW7 2BZ, UK}
\centerline{smolin@ic.ac.uk}
\blankline
\centerline{February 18, 2001}
\blankline
\centerline{ABSTRACT}
A new matrix model is described, based on the exceptional
Jordan algebra, $J^3_{\cal O}$. The action is cubic, as in matrix Chern-Simons
theory.  We describe a compactification that, we argue, reproduces,
at the one loop level,
an octonionic compactification of the matrix string theory in which
$SO(8)$ is broken to $G2$.  There are 27 matrix degrees of
freedom, which under $Spin(8)$ transform as the vector, spinor and
conjugate spinor, plus three singlets, which represent the two longitudinal
coordinates plus an eleventh coordinate.  Supersymmetry appears to
be related to triality of the representations of $Spin(8)$.  
\vfill

${}^*$ smolin@phys.psu.edu
\eject
\tableofcontents
\blankline

\section{Introduction}

This letter proposes a new answer to the question of what
is the background independent
formulation of string theory.  This proposal is based on the use
the exceptional Jordan algebra, whose structure
is shown to code the degrees of freedom of string theory. 
The dynamics of the theory is expressed in a purely algebraic
framework, which is a background independent matrix model, whose
action makes no reference to a metric or classical manifold.
Space and time, as well
as the string and branelike excitations, arise by expanding the
theory around some of its classical solutions.  
 
While this model has a unique algebraic structure, it falls into
a general class of models previously studied called the cubic
matrix models\cite{MCS,LSD,AIKO}, which are supersymmetric
extensions of matrix Chern-Simons theory. These models are of interest
because they appear to  provide a unification
of  string theory with a class of background independent
theories. These latter are  extensions of loop quantum gravity\cite{lqg},
which were previously studied as a possible background
independent framework for $\cal M$ theory\cite{tubes,pqtubes,mpaper}.

Just as string theory appears to be the unique framework for 
constructing quantum gravity in the background dependent regime,
loop quantum gravity\cite{lqg} is so far the only well worked out framework
we have for describing quantum gravity in a background independent
fashion.  The dynamics can be expressed in either a 
hamiltonian\cite{lqg}
or path integral formalism\cite{foam,FM3}, when the latter is used
close relations to topological field theory and dynamical
triangulations are revealed. The framework is quite general,
and accommodates not only quantum general relativity\cite{lqg},
but supergravity\cite{superloop}, branes\cite{loopbranes}
and higher dimensional 
theories\cite{higher,11}. 
The framework also provides a natural formulation of the holographic
principle which makes sense in cosmological and background
independent contexts\cite{weakholo}.  Since the real quantum
theory of gravity must be formulated in a background independent
manner, yet have a consistent description at the background
dependent level, it is reasonable to seek the theory by means of
a unification of loop quantum gravity and string theory. The
search for this unification\cite{tubes,pqtubes,mpaper,stringsas} led
to the discovery of the cubic matrix models, which appear to have
both background dependent phases that reproduce the known
matrix models for string theory\cite{MCS,AIKO} and 
Chern-Simons-like phases that lead to a version of loop
quantum gravity with the symmetries appropriate to string 
theory\cite{LSD}.
 
The next question facing this approach is whether
there may be within it a single theory based on 
a unique mathematical structure which could
serve as a framework for $\cal M$ theory, which is the theory which is
conjectured to unify the different perturbative string theories.  The
mathematical structure should be closely related to the symmetries
of string and $\cal M$ theory, but it should be unique, so that we are no longer 
allowed to ask, for example, why a theory based on $Osp(1|8)$ or
$Osp(1|256)$ is not as good as one based on $Osp(1|32)$?  The unique
structure we seek should also be able to answer in a simple way some
of the most puzzling aspects of string theory, such as 
why is it that the perturbative consistency of a quantum
theory of gravity seems to require $6$ more spatial dimensions than we
observe, or why string theories 
require at least $16$  supersymmetry generators, when none have so
far been observed.  

There is a very elegant answer to both of these questions 
which has been proposed from time to 
time\cite{ostructure,books,CT}, which is that the quantum geometry 
of space must involve
the structures of the octonions.  The idea, roughly, is that
the local\footnote{Because 
global symmetries can play no role in the formulation of a background
independent gravitational theory, classical or quantum.} structure 
of a $9+1$ dimensional spacetime may be
expressed in terms of $J_2^{\cal O}$, which is the Jordan algebra
of $2\times 2$ hermitian matrices of octonions\footnote{For details
about octonions and Jordan algebras, good references are 
\cite{adams,books}.},
\f
J^2_{\cal O} = \left (
\begin{array}{cc}
    z_1 & {\cal O}_0  \\
     \bar{{\cal O}}_0 & z_2  
\end{array} \right )
\ff
It is elementary to show that this is a representation of
$9+1$ dimensional Minkowski spacetime, $M^{9,1}$. 
Here $z_1$ and $z_2$ can be taken as the light cone coordinates
$z_\pm$ and the $8$ transverse coordinates are coded in the
octonion ${\cal O}_0$.  
Similarly,
the relevant $Spin(9,1)$ spinors may be parameterized 
as $2$ component octonionic spinors,
which we may write as,
\f
\Psi = \left (
\begin{array}{c}
   \bar{\cal O}_2  \\
       {\cal O}_1
\end{array} \right )
\ff
Furthermore, several authors have pointed out that the structure
of string theory and ten dimensional super-Yang-Mills theory 
requires identities that have to do with the existence
of the octonions\cite{ostructure,books,CT}.

What would we observe if nature were like this?  The problem is that
it is hard to think of measurements that will produce
octonionic valued quantities: most measurements give back only coarse
grained, averaged quantities, which are usually rational numbers.

However, it is intriguing to wonder whether this may be the
answer to the question we raised. One may conjecture that when
octonionic variables are averaged over in some 
coarse graining procedure, they reduce to complex variables, as 
under the coarse graining the fields may forget the delicate algebraic
relations that underlie the lack of associativity and commutivity of
the octonions.  Further,  commutivity and associativity of measured
values may be forced on us by the procedures by which we measure
spacetime geometry.  If this is the case then under coarse graining
we would observe the fundamental structures reduced by 
a projection map\footnote{This is inspired by a conjecture of
Dray and Magnogue\cite{CT}.}
\f
\Pi: {\cal O} \rightarrow {\bf C}.
\ff
But under this map
\f
J^2_{\cal O} \rightarrow J^2_{\bf C} 
\ff
which is a representation of $3+1$ dimensional 
spacetime, while
the spinors reduce to the two component chiral spinors of
$SL(2,C)$.  

The main idea of supersymmetry, in the context of a theory of 
spacetime, is that there should be some
fundamental unification of the spacetime geometry with the fermionic
degrees of freedom which live in the spinor representations of
the local invariance group.  There has recently been much
speculation about the existence of an $\cal M$ theory, which
unifies the different string theories in the context of an $11$
dimensional structure.  It is then very intriguing to notice
that there is a way to unify the coordinates of
the tangent space of a $9+1$ dimensional spacetime
with its spinor degrees of freedom, in a way that naturally
includes an eleventh spatial coordinate $z_3$.  This is to
incorporate all of them in the algebra of $3 \times 3$
hermitian matrices of octonions, which is called the
exceptional Jordan algebra.  This is given by
\f
J^3_{\cal O}= \left (
\begin{array}{ccc}
    z_1 & {\cal O}_0 & \bar{{\cal O}}_2 \\
     \bar{{\cal O}}_0 & z_2 &  {\cal O}_1 \\
    {\cal O}_2 &   \bar{{\cal O}}_1 & z_0
\end{array} \right )
\ff
We see that the three real and three octonionic variables
gives us $27$ degrees of freedom.  In this letter we will
show that these can parameterize a matrix model for string theory
in which the vector and spinor of $9+1$ dimensional Minkowski
spacetime are unified with an eleventh spatial coordinate.  
At the same time, the $27$
degrees of freedom reminds us of a proposal to formulate
$\cal M$ theory in terms of an extension of bosonic
string theory to a $27$ dimensional theory\cite{garylenny}, as well
as the idea that 
sixteen of the dimensions of bosonic string theory are  transmuted
from bosonic to fermionic by a dynamical mechanism that involves
the decay of the tachyonic degree of freedom.  

It then seems possible that the exceptional Jordan algebra
is exactly the unique mathematical structure we seek.  To
investigate this hypothesis we here formulate a theory
based on this algebra. We find that the exceptional Jordan
algebra has invariants which allow the formulation of
an apparently unique theory, when 
formulated in the language of the cubic matrix model\cite{MCS,LSD}.

We may recall that the basic idea of the cubic
matrix model is to build a background independent
matrix model for string/M theory by starting with a matrix
representation of Chern-Simons theory and then to extend
it by taking degrees of freedom in an algebra associated
with the conjectured tangent space symmetries relevant
for string theory. By expressing the dynamics in terms
of matrix Chern-Simons theory rather than a matrix model
derived from a compactification of Yang-Mills theory, dependence
on a particular background manifold and metric
is avoided.  At the same time, as shown in \cite{LSD}, this makes 
possible compactifications that give rise to completely manifold
independent formulations of the theory\cite{FM3,tubes,pqtubes} 
which are then seen
to be described by a quantum deformation of the kinematical and dynamical 
structures of
loop quantum gravity.
In \cite{MCS,LSD} models were proposed based on $Osp(1|32)$ and
its complexification $SU(16,16|1)$, here a simpler
model of the same kind is proposed based on the 
exceptional Jordan algebra, $J^3_{\cal O}$.  The $Osp(1|32)$ model was
argued in \cite{MCS} to reproduce both the dWHN-BFSS\cite{CH,dWHN,BFSS} and
IKKT\cite{IKKT} models under particular compactifications, once the
one loop corrections to the effective action were taken
into account\footnote{A recent study of these models is 
reported in \cite{AIKO}.}. 
Here we show that a similar argument applied
to  $J^3_{\cal O}$ leads to a certain compactification of 
the matrix string theory
described in \cite{motl,more,dvv} under which $SO(8)$ is broken
to $G2$, the automorphism group of the octonions.  
We caution, however, that
the argument here is based only on constraining the form of the
effective action by its expected symmetries, detailed calculations
based on a BRST quantization of Matrix Chern-Simons theory
given in \cite{brst} are in progress and will be reported
elsewhere. 

Finally, we mention that B. Kim and A. Schwarz have proposed an elegant
formulation of the IKKT model based on $J^2_{\cal O}$ and
its spinor representation\cite{albert}. It is likely that that is related
to a compactification of the model studied here.

\section{The exceptional Jordan algebra}

The exceptional Jordan algebra, $J$ is composed of $3 \times 3$
hermitian matrices of octonions\cite{adams}.
We will write the components as
\f
J= \left (
\begin{array}{ccc}
    z_1 & {\cal O}_0 & \bar{{\cal O}}_2 \\
     \bar{{\cal O}}_0 & z_2 &  {\cal O}_1 \\
    {\cal O}_2 &   \bar{{\cal O}}_1 & z_0
\end{array} \right )
\ff
where $z_a \in R$ and ${\cal O}_a $ are octonions.

The automorphism group of $J$ is known to be $F_4$.
This group has several $Spin(8)$ subgroups, one of
which acts on the components of $J$ in the following
way: the $z_a$ are scalars, ${\cal O}_0, {\cal O}_1, {\cal O}_2$
are, respectively, the $8$ dimensional vector,
spinor and conjugate spinors.  This is also
an extension of this $Spin(8)$ subgroup to a $Spin(9)$ subgroup on which
$({\cal O}_1,{\cal O}_2)$ transforms as the $16$ dimensional
spinor representation.  

We also note that there is an $SO(2)$ subgroup of
$F_4$ such that $(z_1,z_2) = z_I$, with $I=1,2$ \
transform as the vector, while $({\cal O}_1, {\cal O}_2) $
transforms as the spinor. 

Among the automorphisms of $J$ are a discrete set
which we call the triality generators. 
This is an algebra generated by $I$, and $\rho$,  where
\f
\rho \circ J= \left (
\begin{array}{ccc}
    z_2 & {\cal O}_1 & \bar{{\cal O}}_0 \\
     \bar{{\cal O}}_1 & z_0 &  {\cal O}_2 \\
    {\cal O}_0 &   \bar{{\cal O}}_2 & z_1
\end{array} \right )
\label{triality}
\ff
We note that
$(\rho )^3=I$. These operations are called
triality because they mix up the vector, spinor
and conjugate spinor representations of $SO(8)$, and
hence generalize the duality that exchanges the two
spinor representations of even spin groups.

In the following we will make use of the following properties
of the algebra of octonions. An octonion will be
written as ${\cal O}=o_{\vec{a}} e^{\vec{a}}$ with units $e^{\vec{a}}$,
${\vec{a}}=0,1,...,7= (0,i)$. $e^0$ is the identity and the imaginary
units$e^i$, $i=1,...,7$  satisfy
\f
e^i e^j = - \delta^{ij} + \sigma^{ijk} e_k
\ff
where $\sigma^{ijk}$ is completely antisymmetric
and indices are raised and lowered with the flat metric
$\delta_{ab}$ on $R^8$.  The algebra is non-associative and the 
associator is defined as
\f
(e^i e^j ) e^k - e^i ( e^j e^k )= \rho^{ijkl}e_l
\ff
$\rho^{ijkl}$ is also completely antisymmetric and is equal to
\f
\rho^{ijkl} = { 1 \over 3!} \epsilon^{ijklmno} \sigma_{mno}
\ff
The automorphism group of the algebra of octonions is $G2$,
which is a $14$ dimensional subalgebra of $Spin(7)$. 

\section{The model}

We note that the $SO(8)$ representation content of $J$ 
contains the fields of the matrix string model, plus an 
additional scalar degree of freedom.  It is then very suggestive
that this extra matrix degree of freedom, which is
$z_0$, corresponds to the $11$'th dimension  of $\cal M$ theory.
To test this idea we construct and study a matrix model
based on $J$.  

The degrees of freedom of our model will live in 
${\cal G} \times J$, where $\cal G$ is a Lie algebra. In this
letter we take ${\cal G}=U(P)$. A configuration of the
system is then given by $J_I \otimes g^I$, where
$g^I$ are the generators of $\cal G$. We will sometimes
write $J_A^{B} = J_I \otimes g_A^{I \ B}$ where
$g_A^{I \ B}$ are the generators of $U(P)$ in the fundamental,
$P$ dimensional representation.

We want to make an action which is a functional
of the $J_A^{\ B}$. If we follow the strategy of
\cite{MCS,LSD} we may try to make a background independent
model which should contain within it a topological
quantum field theory.  One way to do this is
to construct a cubic action of the kind studied
in \cite{MCS,LSD}.  This means the action should
be antisymmetrized in such a way that it reproduces
the matrix form of Chern-Simons theory with the
three $N\times N$ matrices $z_a$.  

To construct the action we need a cubic product
on $J$. There is a unique cubic product,
\f
t(J_1 , J_2 , J_3 )  \rightarrow R
\ff
which is invariant under $F_4$. It is, however, completely symmetric.

The problem is that to get an action invariant under
$\cal G$ we have to combine $t$ with the structure 
constants of $\cal G$, $f_{IJK}$, which are antisymmetric.  To combine
them  we need another antisymmetrization.  One possibility
is to use the triality map (\ref{triality})
to define an action:
\f
S= {k \over 4\pi} f_{IJK}
t( J^I , \rho \circ J^J , \rho^2 \circ J^K )
\ff
This action defines the theory, which we will call
the {\it exceptional cubic matrix model.}   

Expanded out in its $Spin(8)$ components, the
action reads
\f
S={k \over 4\pi}f_{IJK}\epsilon^{abc } \left \{
3 x_a^I x_b^J x_c^K 
+9 x^K_c Re (\bar{{\cal O}}_a^I {\cal O}_b^J )
-3 Re \left ( ({\cal O}_a^I {\cal O}_b^J ) {\cal O}_c^K \right )
\right \} + 
\sum_a f_{IJK}\sigma^{ijk} {\cal O}_{a i}^I {\cal O}_{a j}^J
{\cal O}_{a k}^K
\ff
where $x_0 = z_1 + z_2$ and cyclic. We note that 
$Re[(O_1 O_2) O_3]=f(O_1, O_2, O_3)$, the triality map
defined in \cite{adams} and 
$Re(\bar{O}_1 O_2) = (O_1 , O_2)$ the inner product on the
$8$ dimensional space of octonions.  

The first term will give rise to matrix chern-simons 
theory\cite{MCS,LSD}.
This is the reason the coupling is written as ${k \over 4\pi}$.
For the compactified action to be well defined under
global gauge transformations, $k$ must then be an integer.  

The action has several symmetries. These include
$Spin(8) \otimes U(P)$ transformations, triality
and matrix translations,
\f
J_{A}^{\ B} \rightarrow J_{A}^{\ B} + j \delta_A^{\ B}
\ff
where $j$ is a member of the exceptional jordan algebra.
We note that $Spin(8)$ acts differently on each
of the three slots of the cubic product. 
The action is also invariant under $G2$ transformations
which is the automorphism group which leaves the
coefficients $\sigma^{ijk}$ invariant. 
The three particular
$Spin(8)$ subalgebras that act on each slot are related to
each other by triality.

\section{Three torus compactification}

We now consider a standard matrix 
compactification\cite{wati,MCS,LSD} to a three-torus, on
the three coordinates $x_a$.  We break the $U(P)$ indices into 
indices $P,Q=1,...,N$ and $i_0, i_1,i_2$ such that
$i_a=1,...M_a$ and $P=M_0 M_1 M_2 N$, We then write as usual
\f
x_{a Pi_a}^{Qj_a} = {\partial_a}_{i_a}^{j_a} I_P^Q + a_{ai_aP}^{j_aQ}
\ff
where the gauge fields $a_{ai_aP}^{j_aQ}$ are compactified
as bosonic fields so that
\f
a_{a(i_0+M_0i_1 i_2)P}^{(j_0+M_0 j_1 j_2) Q}=a_{a(i_a)P}^{(j_a) Q}
\ff
We similarly compactify ${\cal O}_0$.
However, ${\cal O}_1,{\cal O}_2$ are compactified as fermions, 
since they live in the spinor representations of
$Spin(8) \oplus SO(2)$, so that
\f
{\cal O}_{A(i+M_0)P}^{(j+M_0) Q}=-{\cal O}_{A(i)P}^{(j) Q}
\ff
for $SO(2)$ spinor indices $A=1,2$.  We compactify with the
same signs in the $x^1$ and $x^2$ directions.

The resulting action can be approximated by a continuum action for
very large values of the $M_a$. We find
\begin{eqnarray}
S& = &{k \over 4\pi L_0L_1L_2} \oint_{T^3} d^3x  Tr_N \left \{  
 [\epsilon^{abc} ( a_a \partial_b a_c + {2\over 3} a_a a_b a_c )
+ \epsilon^{AB} Re (\bar{{\cal O}}_A {\cal D }_0 {\cal O}_B )  \right \}
\nonumber \\
&&+ 
f( {\cal O}_{0 P}^{\ Q} ,  
{\cal O}_{1 Q}^{\ R},{\cal O}_{2R}^{\ P}  ) + 
f_{IJK}\sigma^{ijk} {\cal O}_{0 i}^I {\cal O}_{0 j}^J {\cal O}_{0 k}^K
\end{eqnarray}
where $f({\cal O}_0, {\cal O}_1 , {\cal O}_2 )$ is the
triality map\cite{adams}.
Here the compactification radii
may be expressed as $L_a = l_{Planck} M_a$ where a
dimensional scale is introduced for convenience and
taken to be $l_{Planck}$. As in 
\cite{MCS,LSD} the physics is assumed to imply that
the $M_a$ are very big, but not infinite and that $l_{Planck}$
denotes a fixed physical scale, which as in loop quantum gravity
is related to minimal areas and volumes.  Hence the model
is finite and all quantum corrections are finite, although there
may be infrared divergences as the $M_a$ diverge.

We may note that there is a possible term which is  linear
in fermionic variables, of the form
\f
( {\cal O}_{0 P}^{\ Q} , 
{\cal D}_{1Q}^{\ R} {\cal O}_{2 R}^{\ P})  -
( {\cal O}_{0 P}^{\ Q} , 
{\cal D}_{2Q}^{\ R} {\cal O}_{1 R}^{\ P}).
\ff
However that term is inconsistent with the boundary conditions
as a translation by $M_0$ takes it to minus itself, hence
it vanishes and does not appear in the compactified action. 
Other terms of the form
\f
\sum_{A=1}^2 f_{IJK}\sigma^{ijk} {\cal O}_{A i}^I {\cal O}_{A j}^J {\cal 
O}_{A k}^K
\ff
vanish once the spinorial variables ${\cal O}_{A j}^J$ are
taken to be fermions, by the combination of the three antisymmetries.

As a result, triality and $SO(2,1)$ are broken in the action, by
the fermionic boundary conditions. But the action
still has local $U(N)$ gauge symmetry, global $G2$
symmetry and matrix translation symmetry in the variables
\f
a_{aP}^{\ Q}(x) \rightarrow a_{aP}^{\ Q}(x) + v_a \delta_P^{\ Q}
\ff
\f
{\cal O}_{aP}^{\ Q}(x) \rightarrow {\cal O}_{aP}^{\ Q}(x) + o_a \delta_P^{\ Q}
\ff
where $o_a$ is a triplet of octonions.

\section{The one loop action and matrix string theory}

The next step is computation of the effective action. While this
has not yet been done we may use symmetries and power counting to constrain
the form of the effective potential at one loop. At the one
loop level we expect to have all terms of dimension $4$ consistent
with the symmetries. This gives us terms of the following forms
(with $a=(0,x)$, and $x,y,=1,2$ the $2d$ spatial coordinates),
\begin{eqnarray}
I^{\hbar} &=&  {1 \over L_0 L_1 L_2} \oint_{T^3} d^3x  Tr_N   \{ 
f_{xy}f^{xy} + f_{0x}f^{0x}
+ ([{\cal D}_x , {\cal O}_0 ],[{\cal D}^x , {\cal O}_0])   
 \\
&&   + ([{\cal D}_0 , {\cal O}_0 ],[{\cal D}^0 , {\cal O}_0])
+ \tau^{xAB}({\cal O}_A , [{\cal D}_x, {\cal O}_B )
+ [ {\cal O}_{0i} , {\cal O}_{0j} ][ {\cal O}_{0k}  , {\cal O}_{0l}  ]
(\alpha \rho^{ijkl} + \beta \sigma^{ijn}\sigma^{kl}_{\ \ n} )
  \} \nonumber
\end{eqnarray}
where $\tau^{xAB}$ is the two dimensional Pauli matrix. 
We want to emphasize that from the symmetry analysis we cannot
fix the coefficients, but only the terms.  

To get to a theory related to the matrix 
string we  take the limit $L_0 \rightarrow L_{Planck}$ so there is only the
lowest mode in the $0$ direction. We then drop terms in
$\partial_0$. For consistency with the gauge invariance we must
drop terms in $a_{0P}^{\ Q}$ at the same order. 

The result is a two dimensional effective action which is of the form
\begin{eqnarray}
I^{eff}&=&I^0 +I^\hbar = \oint_{T^2}d^2x  Tr_N
 \{  f_{xy} f^{xy} + 
({\cal D}_x {\cal O}_0 , {\cal D}^x {\cal O}_0 )
+\tau^{x AB}({\cal O }_A , [{\cal D}_x , {\cal O}_B ])
\nonumber \\
&&+f({\cal O }_0, {\cal O }_1 , {\cal O }_2 )
+  [ {\cal O}_{0i} , {\cal O}_{0j} ][ {\cal O}_{0k}  , {\cal O}_{0l}  ]
(\alpha \rho^{ijkl} + \beta \sigma^{ijn}\sigma^{kl}_{\ \ n} )
 \}
\end{eqnarray}
Using the triality, we may write the 8 dimensional vector 
representation as ${\cal O }_0= V_{\vec{a}} e^{\vec{a}}$, 
where $\vec{a}=0,...,7$
and $e^{\vec{a}}$ are the generators of the octonions.  Given $Spin(8)$
spinor indices $\alpha=1,...,8$ and $\bar{\alpha}=\bar{1},...,\bar{8}$
we then write ${\cal O}_1=S_\alpha e^\alpha$ and
${\cal O}_2=\bar{S}_{\bar{\alpha}} e^{\bar{\alpha}}$.  We then have,
\begin{eqnarray}
I^{eff}&=&  \oint_{T^2}d^2x  Tr_N
 \{  f_{12} f^{12} + 
({\cal D}_I {V}_{\vec{a}}  {\cal D}^I {V}^{\vec{a}} 
+\sigma^{I AB}(S_A , [{\cal D}_I , S_B ])
+ \Gamma^{\vec{a}\bar{\alpha} \alpha }V_{\vec{a}} , 
\bar{S}_{\bar{\alpha}} S_\alpha 
\nonumber \\
&&+[ V_{i} , V_{j} ][ V_{k}  , V_{l}  ]
(\alpha \rho^{ijkl} + \beta \sigma^{ijn}\sigma^{kl}_{\ \ n} ) \}
\end{eqnarray}
This is model has the degrees of freedom of the matrix
string theory described in \cite{dvv,motl,more}.
The only difference between this action and that of the
matrix string is in the form of the four-matrix
interaction terms among the $V_{\vec{a}}$, which reflect
the fact that the $SO(8)$ symmetry of the bosonic fields
of the matrix string has been broken to $G2$.

We may note that with particular coefficients on the cubic
and quartic terms, the matrix string action is supersymmetric.
It is interesting to conjecture that the supersymmetries are 
preserved under the breaking of $Spin(8)$ to $G2$ and that it can
be understood as arising
from the components of $F_4$ that live in the
$8$ dimensional spinor and conjugate spinor representations
$S$ and $S_c$.  These act bosonically on the degrees of freedom
in $J^3_{\cal O}$ however it is possible that after the
compactifications that turn the ${\cal O}_A$ into fermions
they may imply a fermionic symmetry of the reduced action.
This is currently under investigation.  In this regard
we may note that manifolds with $G2$ holonomy have been
studied before in connection with compactifications of
supergravity and string theory\cite{G2}.

\section{Closing comments}

In this note we have only introduced the cubic matrix model
based on the exceptional Jordan algebra; many things need
to be investigated. The first is to have some understanding of
whether the quantization of the spinorial degrees of freedom
as fermions is a free choice or is forced on us by something
like a spin statistics theorem. Related to this is the question 
of whether the model is supersymmetric after the compactification
proposed here, and whether the supersymmetries can be understood as
related to that part of the $F_4$ algebra that is generated by
$Spin(8)$ spinorial variables. It is also interesting to wonder
whether the $27$ components of $J^3_{\cal O}$ are the same
$27$ dimensions that have been recently proposed for
a bosonic form of $\cal M$ theory\cite{garylenny}. 
Finally, it is very interesting to
investigate whether compactifications such as described in
\cite{LSD} are allowed, as these will give a background
independent phase which can be described in the language of
quantum deformed loop quantum gravity, and thus show that there
is a form of loop quantum gravity dual to the matrix string theory.

\section*{ACKNOWELDGEMENTS}

I am first of all very grateful for many discussions with
Richard Livine and Yi Ling during the course of this work as
well as to Arivand Asok for a collaboration two years ago
on these topics. 
I am also grateful to conversations with John Baez, 
Martin Cederwall, Louis Crane, Tevian Dray, Murat Gunyadin,
Corinne Magnogue, Fotini Markopoulou and Albert Schwarz about octonions and
their possible role in physics.  Comments from John Baez, Clifford
Johnson, Richard 
Livine and Albert Schwarz on a draft of this paper were very helpful.
I am also grateful to Chris Hull
for asking if there is a form of Matrix Chern-Simons theory
that has a unique structure related to string theory.
This work was supported by the NSF through grant
PHY95-14240 and gifts from the Jesse 
Phillips Foundation.

\end{document}